\documentclass[10pt,conference]{IEEEtran}
\usepackage[utf8]{inputenc}
\usepackage[T1]{fontenc}
\usepackage{microtype}
\usepackage{rotating}
\usepackage{caption}
\usepackage{booktabs} 
\usepackage{hyperref}
\usepackage{xcolor}
\usepackage{xcolor}
\usepackage{balance}
\definecolor{verde}{rgb}{0.25,0.5,0.35}
\definecolor{jpurple}{rgb}{0.5,0,0.35}
\definecolor{darkgreen}{rgb}{0.0, 0.2, 0.13}

\usepackage{listings}
\usepackage{pxfonts}

\newcommand{\codeJavascript}{
\lstset{
    language=Java,
    basicstyle=\ttfamily\small,
    keywordstyle=\color{jpurple}\bfseries,
    stringstyle=\color{red},
    commentstyle=\color{verde},
    morecomment=[s][\color{blue}]{/**}{*/},
    morekeywords={function},
    extendedchars=true,
    showspaces=false,
    showstringspaces=false,
    breaklines=true,
    backgroundcolor=\color{white},
    breakautoindent=true,
    captionpos=b,
    xleftmargin=0pt,
    tabsize=2
}}

\begin{document}
\title{An Exploratory Study of Code Smells in Web Games}

\author{
\IEEEauthorblockN{Vartika Agrahari, Sridhar Chimalakonda}
\IEEEauthorblockA{Research in Intelligent Software \& Human Analytics (RISHA) Lab\\Indian Institute of Technology Tirupati India \\
cs18m016@iittp.ac.in, ch@iittp.ac.in}
}

\maketitle
\begin{abstract}

With the continuous growth of the internet market, games are becoming more and more popular worldwide. However, increased market competition for game demands developers to write more efficient games in terms of performance, security, and maintenance. The continuous evolution of software systems and its increasing complexity may result in bad design decisions. Researchers analyzed cognitive, behavioral and social effects of games. Also, gameplay and game mechanics have been a research area to enhance game playing, but to the extent of our knowledge, there hardly exists any research work that studies the bad coding practices in game development. Hence, through our study, we try to analyze and identify the presence of bad coding practices called code smells that may cause quality issues in games. To accomplish this, we created a dataset of 361 web games written in \textit{JavaScript}. On this dataset, we run a \textit{JavaScript} code smell detection tool \textit{JSNose} to find the occurrence and distribution of code smell in web games. Further, we did a manual study on 9 web games to find violation of existing game programming patterns. Our results show that existing tools are mostly language-specific and are not enough in the context of games as they were not able to detect the anti-patterns or bad coding practices which are game-specific, motivating the need of game-specific code smell detection tools.

\end{abstract}


\maketitle


\section{Introduction}
In the past few decades, the game industry has emerged as one of the significant contributors to the world's economy \cite{marchand2013value}. Because of its growing popularity and increased market demand,
the complexity of these games is also increasing.
Thus, it becomes crucial and necessary for researchers to analyze games from a quality perspective so as to guide developers in making efficient and quality games \cite{murphy2014cowboys}. One way to ensure the quality of game development is to minimize the inadequate practices used to write them called \textit{code smells}.
Code smells are bad design choices or bad coding practices incurred at the time of development. However, they can be taken as an analogy to patterns in software programs which are related to bad designs \cite{van2002java}. Code smells are not bugs and their presence in the code does not make code to deviate from the expected execution. Nevertheless, their presence may lead to long-term maintainability problems and technical debt \cite{fowler1997refactoring, mantyla2003taxonomy}.

Several research studies and technical works have been done to study the presence of code smells in different domains \cite{mantyla2003taxonomy, sharma2018survey, fernandes2016review}. Fowler proposed the need to identify code smells \cite{fowler1997refactoring}, following which researchers proposed various rules and standards to detect and refactor them \cite{mantyla2003taxonomy, fard2013jsnose}. Numerous tools have been proposed to detect code smells in software applications by many developers and researchers \cite{fernandes2016review, nguyen2012detection,fard2013jsnose}.
However, understanding and analyzing the presence of code smells in games is still largely unexplored in the literature motivating the need for our work.

Game development is different from other domains of software development because games involve rendering process, real-time constraints, AI and physics components and many more performance-related factors which plays an important role in process of game development and may not be specific to other domains \cite{murphy2014cowboys}. Presence of code smells in games can affect the process of development adversely and ultimately leads to future issues of performance, security, and maintainability \cite{ghafari2017security, sjoberg2013quantifying, yamashita2012code, stamelos2002code}. Hence, it becomes critical to identify and discard these smells from the games to ensure performance and maintainability. Language-specific code smell detection tool may not be sufficient to detect all types of code smells present in games, thus we did this work to investigate the presence of existing code smells in games. To the best of our knowledge, existing literature on code smells did not focus on the domain of web games, thus motivating the need for our study, although a brief work done on code smells in games by Vaishali \cite{khanve2019existing} emphasized the relevancy of existing code smells in games. Even though resources to identify code smells in games are limited, there is a need to investigate the presence of existing language-based code smells in games to understand the frequency and distribution of these bad smells. 

In this study, we make an attempt to analyze the occurrence and distribution of code smells in web games using a well-known existing code smell detection tool \textit{JSNose} \cite{fard2013jsnose} on our dataset, which consists of 361 \textit{JavaScript} games mined from Github. Furthermore, as a preliminary work, we did a manual study on 9 of these 361 \textit{JavaScript} games to find the violation of game programming patterns in games that could be detected by existing code smell detection tools. We conducted a preliminary study to find the relevance of code smells in web games. Our study answers the following research questions:
\begin{itemize}
    \item \textbf{RQ1:} \textit{What is the distribution of existing code smells in the context of games?}
    \item \textbf{RQ2:} \textit{If game-specific code smells exists, how they are different from existing ones?}
    \end{itemize}
The remainder of the paper is structured as follows. Section \ref{motivation} discusses the research motivation behind this work. To set the context, Section \ref{jsnosesmells} presents a list of existing code smells in \textit{JavaScript} programs along with a brief introduction to each of them. Section \ref{gamepatterns} discusses the violation of game programming patterns followed by Section \ref{empirical} which focuses on the methodology of empirical study being done. Section \ref{relatedwork} presents the literature review of the work followed by limitations in Section \ref{discussion}. Conclusion and Future work are discussed in Section \ref{conclusion} and Section \ref{future} respectively.


\section{Research Motivation}
\label{motivation}
An empirical study by Guo et al. states that code smells can vary from domain to domain \cite{guo2010domain}. A code smell which has a serious impact on one domain may be less effective in other domain. Thus, code smells cannot be generalized to be language-specific. A study done on the performance impacts of code smells in Android applications shows that refactoring brought considerable improvement in code quality \cite{hecht2016empirical}. Similar studies on code smells in domains like software product lines \cite{abilio2015detecting} and spreadsheet formulas \cite{hermans2012detecting} resulted in an appreciable gain in terms of performance and business processes respectively. Thus, we believe that code smells should be considered in the case of games as well, to improve the code quality and performance of games. Researchers studied the psychological and social \cite{lee2006we}, educational \cite{giessen2015serious} and behavioural \cite{anderson2001effects} effects of games. Also, gameplay and game mechanics are discussed by researchers \cite{fabricatore2007gameplay} but to the extent of our knowledge, hardly any study exists which studies games from the quality perspective.
Expectations on games are getting increased and quality of games, game development process is getting complex and developers are inlcuding more and more features to make games more adventurous and fun to play. This inclusion of enormous features and disregard of future consequences may lead to faulty games.

Game development differs from the traditional software development in the context of different performance and real-time factors being involved like memory allocation and deallocation issues, rendering process and graphical components \cite{murphy2014cowboys, kasurinen2013game, nystrom2014game}. Thus, language-specific code smell detection tools may not be sufficient to handle code smells of games. To strengthen our research motivation, our goal is to investigate the scope of code smells in games and to verify the necessity of game-specific code smell detection tool.

We choose web games because they are accessible, affordable, easy to play and does not require any installation or dependencies. They are generally preferred by end-user over games with dependencies. Web games are mainly developed in \textit{JavaScript, HTML}, and \textit{CSS}, and thus we choose open-source JavaScript games to conduct our empirical study. Also, \textit{JavaScript} is an evolving object-based scripting language, thus code smells are mainly dependent on identifying objects, functions, and classes \cite{fard2013jsnose}. Moreover, a study shows that \textit{JavaScript} is still an unexplored language, and needs to be considered, as the number of \textit{JavaScript} projects is increasing \cite{kostanjevec2017preliminary}. 

Since we chose web games for the exploration of code smells, we need existing code smells detection tools in \textit{JavaScript}. A study \cite{sharma2018survey} shows that two \textit{JavaScript} code smells detection tool exist named as \textit{JSNose} \cite{fard2013jsnose} and \textit{JSpIRIT} \cite{vidal2015jspirit}. We observed that the smells detected by \textit{JSNose} is almost superset of the smells detected by \textit{JSpIRIT}. Also, we find that \textit{JSNose} has been verified by researchers giving good precision score \cite{rasool2015review}.  Thus, we chose \textit{JSNose} for our empirical study. 

We aimed at identifying the violation of game programming patterns. If such violation happens, code smells in games need to be handled separately supported by game-specific code smells detection tool.




 

\begin{table*}[t]
  \centering
  \caption{Meta information of 9 games selected for manual study}
  \begin{tabular}{|c|c|c|c|c|c|l|}
  \hline
    \textbf{Game Name}&\textbf{Category} & \begin{turn}{90}\textbf{\#JS Files}\end{turn} & \begin{turn}{90}\textbf{\# JS LOC} \end{turn}&\begin{turn}{90} \textbf{\#Stars}\end{turn} & \begin{turn}{90}\textbf{\#Issues}\end{turn}& 
    \textbf{Link}\\
    \hline
    \hline
    Cube-composer & Puzzle & 6 & 230 & 1389& 8&\url{https://github.com/sharkdp/cube-composer}\\
    
     \hline
       Clumsy-bird &Arcade &7 & 571& 1210&3 &\url{https://github.com/ellisonleao/clumsy-bird}\\
    \hline
    Javascript\_snake &Arcade&5&846&3&1&\url{https://github.com/gamedolphin/javascript_snake}\\
    \hline
     Diablo-js & Roleplaying game(RPG)& 1&743&767&5&\url{https://github.com/mitallast/diablo-js}\\
    \hline
    Digger & Arcade & 2&971&39&1&\url{https://github.com/lutzroeder/digger}\\
    \hline
     3D-Hartwig-chess-set &Board Game  & 4& 1511 & 322 & 5 & \url{https://github.com/juliangarnier/3D-Hartwig-chess-set}\\
    
    \hline
    Astray & Maze Game&5&1523&290&5&\url{https://github.com/wwwtyro/Astray}\\
    \hline
    Hextris & Puzzle & 15 & 1935 & 1478 & 16 & \url{https://github.com/Hextris/hextris}\\
    \hline
    Follow ME! A Simon Clone &Board Game&4&345&4&1&https://github.com/gamedolphin/\\
    &&&&&&follow\_me\_javascript\_simon\_clone\\
    \hline
  \end{tabular}
  \label{meta}
\end{table*}


\begin{table*}[t]
  \centering
  \caption{Result of \textit{JSNose} on Web Games}
  \begin{tabular}{|c|c|c|c|c|c|c|c|c|c|c|c|c|c|}
  \hline
     \textbf{Statistics} & \textbf{(s1)} & \textbf{(s2)} &
     \textbf{(s3)} &
     \textbf{(s4)} &
     \textbf{(s5)} &
     \textbf{(s6)} &
     \textbf{(s7)} &
     \textbf{(s8)} &
     \textbf{(s9)} &
     \textbf{(s10)} &
     \textbf{(s11)} &
     \textbf{(s12)} &
     \textbf{(s13)}\\
    \hline
    \hline
     Number of smells & 256687	&
    247	&1361&	8737&	0&	471533&	365789&	51816&	39028&	6233&	88606&	5885&	70\\
    \hline
    Average smell in each game& 711.04	&0.68	&3.77	&24.20	&0	&1306.18&	1013.26	&143.53	&108.11	&17.26	&245.44	&16.30&	0.19\\
    
    \hline
    \% out of all smells & 19.80&	0.01&	0.1	&0.67&	0&	36.3&	28.22&	3.99& 3.01	&0.48&	6.83&	0.45&	0.005 \\
    \hline
     \% of games containing smell  & 34.97&	3.82&	11.47&	98.63&	0&	42.62&	31.96&	33.06&	26.90&	19.29&	25.54&	20.38&	1.90\\
    \hline
  \end{tabular}
  
  \label{jsnoseResults}
\end{table*}
\section{Existing Code Smells}
\label{jsnosesmells}
 During software maintenance and evolution, a software system undergoes several levels of changes within restricted time, which ultimately leads to bad design decisions. These poor design choices are code smells \cite{fard2013jsnose}. In software programming, a code smell is any characteristic in the source code of a program that possibly indicates a deeper future problem related to performance, security and maintenance issues. There are different kinds of code smells identified for the language chosen or paradigm used. Our study focuses on code smells present in web games which can be identified using an existing tool called \textit{JSNose}, a \textit{JavaScript} code smell detection tools. According to existing studies \cite{kerievsky2005refactoring, crockford2008javascript}, \textit{JavaScript} includes 7 generic code smells. Later, Fard et al. proposed 6 more code smells during their study. Thus, overall they proposed a code smell detection tools which detects 13 code smells described below.
 \begin{itemize}
\item \textit{Closure Smells (s1)}: This code smell provide inner characteristics of \textit{JavaScript} programming such as over complications arises due to multiple nested functions, accessibility of outer function, variables from the inner function, name conflict arises in the scope of closure, conflict between 'this' assignment to the inner and outer function.

\item \textit{Coupling \textit{JS/HTML/CSS} (s2)}:Coupling of \textit{JavaScript} with \textit{HTML} and \textit{CSS} has been  categorized into 3 code smells. These code smells are \textit{JavaScript} in \textit{HTML}, \textit{HTML} in \textit{JavaScript},and  \textit{CSS} in \textit{JavaScript}. These code smells complicate debugging and software evolution in \textit{JavaScript} applications.

\item \textit{Empty catch (s3)}: During exception handling, whenever the catch block contains zero lines of code, it leads to empty catch code smell. It should be avoided as it causes exception to meet with no response or prevention.
\item \textit{Excessive global variables (s4)}:
 The presence of a large number of global variables in \textit{JavaScript} application leads to more dependent existing modules and because of increased dependency error proneness, and maintainability efforts increases.
\item \textit{Large object (s5)}: Object which is heavily loaded with numerous tasks needs to be refactored and divided into modules. This causes balanced task division among different modules.
\item \textit{Lazy object (s6)}: Objects which are not doing enough work need to be merged or combined with other smaller objects to increase productivity. Existence of lazy objects causes unnecessary overhead as it does less work and occupies resources.
\item \textit{Long message chain (s7)}: Long message chaining leads to complex flow of program which is difficult to understand. It makes the code tough to follow and maintain. 
\item \textit{Long method/function (s8)}: Method with too many lines of code is considered as code smell as it obstructs readability and maintenance.
\item \textit{Long parameter list (s9)}: When the number of parameters is too many, it makes the program harder to comprehend and maintain. 
\item \textit{Nested callback (s10)}: A callback is an argument passed to another parent function. Excessive nested callbacks are harder to comprehend and difficult to maintain because of their asynchronous nature. 
\item \textit{Refused bequest (s11)}: It is a soft code smells which is caused when a \textit{JavaScript} object does not overrides/utilizes most of the properties it inherits. 
\item \textit{Switch statement (s12)}: It often causes redundant code and makes code difficult to maintain. It causes modification of existing code, whenever a new switch is added.
\item \textit{Unused/dead code (s13)}: Dead code is the unnecessary code which gets executed during the run time but is never used. They make program code unnecessarily harder to comprehend. 
 \end{itemize}

\begin{figure*}[ht!]
    \centering
    \includegraphics[width=16cm, height=5cm]{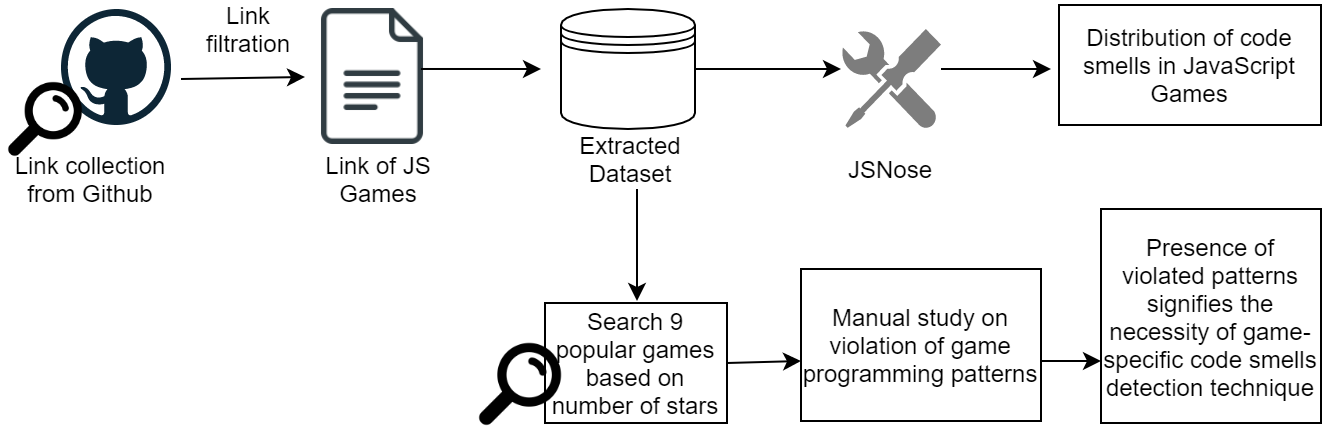}
    \caption{Approach diagram}
    \label{fig:process}
\end{figure*}
\section{Violated Game Programming Patterns}
\label{gamepatterns}
Performance is a critical issue in games, which needs to be met for any game to be appreciated by end-users \cite{klimmt2009player}. Performance comes by good design patterns \cite{bjork2006games} which will optimize the code to an optimal level with better throughput. Although ideally, games should follow numerous patterns \cite{nystrom2014game}, we primarily focused on decoupling and optimization patterns. As an initial experiment, we chose these two patterns because we wanted to see whether such basic design patterns when violated are considered by smells detection tool or not. Decoupling and Optimization patterns should be followed by games as their violation leads to critical issues. Violation of decoupling patterns results in data loss, maintainability and reusability problem, bad user experience on event input, and idle process threads. Whereas optimization patterns violation leads to cache misses, memory fragmentation, deaccelerate memory access, and bad game performance. Thus, considering the importance of these patterns, for the preliminary study we focused on these two patterns. Below we discuss both the patterns with their sub-patterns which play an important role in context of games.
\begin{itemize}
    \item \textbf{Decoupling patterns:} 
    Two modules are said to be decoupled when any changes made in one module does not require any corresponding change in the other module. Thus, this increases the flexibility in software and enhance a better software development process.  Violation of decoupling game programming pattern does not correspond to \textit{Coupling} code smell detected by \textit{JSNose}. As coupling code smell detected by \textit{JSNose} states about occurrence of \textit{CSS} units in \textit{JavaScript}, \textit{JavaScript} in \textit{HTML}, and \textit{HTML} in \textit{JavaScript}. 
    \begin{itemize}
        \item \textbf{Component Decoupling (P1):}
        Game programming includes the creation of different game components such as physics, graphics, sounds, artificial intelligence, simulation, and many more. 
        These game programming components have to be placed separately so that individual can modify single component independently without knowing about other ones. Hence, this makes game development easier and flexible and addresses various maintainability related issues.

        \begin{table*}[t]
  \centering
    \caption{Manual Study of 9 \textit{JavaScript} games}
        \begin{tabular}{|c|c|c|c|c|c|c|c|c|c|c|c|c|c|c|c|}
        \hline
            \textbf{Game} & 
            \textbf{S1} &
            \textbf{S2} & 
            \textbf{S3 } & 
            \textbf{S4} & 
            \textbf{S5} & 
            \textbf{S6} &
            \textbf{S7} &
            \textbf{S8} &
            \textbf{S9} & 
            \textbf{S10} & 
            \textbf{S11} &
            \textbf{S12} &
            \textbf{S13} & 
            \textbf{Violated Patterns}\\
        \hline
        \hline
        Cube-composer &0&	0&	0&	12&	0&	0&	0&	0&	0&	0&0	&0&	0& P3,P2\\
        \hline
         Clumsy-bird & 2&	0&	0&	17&	0&	5&	2&	0&	2&	0&	2&	0&	0& P3, P1\\
        \hline
        Javascript\_snake & 439&	0&	6&	13&	0&	506&	272&	6&	40&	2&	48&	26&	0& P3\\
        \hline
        Diablo-js &70&	0&	0&	11&	0&	90&	4&	4&	0&	0&	14&	2	&0& P1, P3, P4\\
        \hline
        Digger &52&	0&	0&	16&	0	&102&	10&	22&	8&	2&	16&	2	&0& P1, P3\\
        \hline
        3D-Hartwig-chess-set&5&	0&	2&	54&	0	&6&	5&	0&	5&	5&	2&	2&	0&P4, P3\\
        \hline
        Astray &275&	0&	0&	46&	0&	1410&	1524&	28&	234&	0	&970&	70&	0& P4\\
        \hline
        Hextris &250	&0	&88	&9	&0	&835&	6	&94&	46&	0	&30&	9	&0& P4, P3\\
        \hline
          Follow ME! A Simon Clone&323	&0	&4&	15&	0&	301&	195	&4	&38&	2&	48&	8&	0& P1, P3, P4\\
        \hline
    \end{tabular}
  \label{manualstudy}
\end{table*}

        \item \textbf{Event Queue Decoupling (P2):}
        Processing of one event blocks the other event, in this situation other event data might get lost. And if one event took some time to process completely and other events who are blocked (e.g. create sound) now gets the system, causes a delay in request and inappropriate results (in creating sound).
        The global variable is a very common example of an event queue, which is used as a common data element between different components of games. 
    \end{itemize}
\item \textbf{Optimization patterns:} Optimization is the master key to increase the performance of any software entity. Selection of the best possible alternative rewards the developer with better speed and execution. Thus, games are dependent on this factor of optimization to a large extent. Different ways to introduce optimization in game programming pattern are:
\begin{itemize}
    \item \textbf{Data Locality (P3):}
    Arranging data in an order to take advantage of CPU caching is termed as Data Locality. This accelerates the memory access time resulting in less time lag and increased performance.

    Different optimization techniques exist which focuses on providing Data Locality like slicing the data structure, and "hot" and "cold" split. \cite{nystrom2014game}.
    \item \textbf{Object Pool (P4):} While the game is running, we need to create and destroy certain components caused by allocating and freeing memory frequently. This causes the problem of memory fragmentation\cite{nystrom2014game}. To avoid this, we should make an object pool from where we can reuse the objects instead of allocating and freeing them individually. Whenever any object is required it should be searched in the fixed pool, if found, it will be set to be in use, and after freeing of the object, again it will be returned to the object pool. This improves the performance of games, offers better memory management, and utilizes fewer resources.
\end{itemize}
\end{itemize}

\section{Empirical Evaluation}
\label{empirical}
The goal of our study is to find the relevance of existing code smells in the context of games. Additionally, we also aim to find the violation of game programming patterns as they instigate the existence of game-specific code smells. To accomplish the above, we mined 361 games from Github and performed a study on the same to answer the research questions we asked.

\subsection{Dataset creation}
The process of collection of dataset was carried by applying certain filters to make sure we collect the clean data.  
\begin{itemize}
\item As a first step, we made a search on Github with keyword "web games".
\item We selected the language as \textit{JavaScript}.
\item Now simply cloning these games were putting us in suspicion of the false dataset as README of many of the repositories were written in different language. Also, sometimes instead of games, Github gives search result of game-engine. As an instance, we got "Effect-Games"\footnote{https://github.com/jhuckaby/Effect-Games} as one of the search result which is \textit{JavaScript} Game-Engine and Web-Based IDE and does not fall in our research domain.
\item To avoid collecting false dataset, we manually checked README (in English language and containing \textit{JavaScript} game) of each repository, to make sure that our dataset contains \textit{JavaScript} games and nothing else.
\item Thus, following the above process, we collected 361 \textit{JavaScript} games mined from Github.

\end{itemize}

Figure \ref{fig:process} depicts the procedure we followed to conduct this empirical study.
\subsection{Experimental Objects}
To analyze the occurrence and distribution of code smells in \textit{JavaScript} games, we used an existing tool \textit{JSNose}. We mined 361 \textit{JavaScript} games from Github to conduct this experiment. Furthermore, we did a manual study on 9 \textit{JavaScript} games to observe if they follow the game programming patterns or not. The selection of the games was decided based on the number of stars \cite{ray2014large} and \textit{JavaScript} LOC (Lines of Code). While selecting games for manual study, we focused on the fact that the games are popular enough in the developer community and the code size is also significant enough, i.e. neither LOC is too small nor too large so that it is feasible for manual study. To ensure the quality of study with manual feasibility, we chose 9 games for analyzing the violation of patterns. 
The objective of this manual study is to find the violation of game programming patterns that are not detected by existing code smell detection tools. We calculated lines of code using CLOC\footnote{https://sourceforge.net/projects/cloc/}. The average LOC of the 9 web games we studied is 963.88, whereas the average number of stars for 9 game projects is 611.33.
To conduct the manual study, one researcher analyzed the violation of game programming patterns and the second researcher verified each instance of violation. The overall process of manual study took around 60 hours of deep inspection of 9 web games. To maintain the consistency of the result we found, we decided upon certain rules discussed in \cite{nystrom2014game} mentioned in Table \ref{rules}.
Some important statistics of the meta-information of the 9 games have been mentioned in Table \ref{meta}. 

\begin{table*}[t]
\caption{Rules to detect the violation of game programming patterns}
  \centering
  \begin{tabular}{|c|c|c|}
  \hline
  \textbf{SN}&
       \textbf{Game programming pattern} & \textbf{Rules to detect violation of patterns}\\
    \hline
    \hline
    1 & Component Decoupling & Classes and Methods touching multiple components of game, \\  & &or, Monolithic class or, Large Methods\\
    \hline
   2 & Event Queue Decoupling & Input events or events from methods, not stored in central queue\\
    \hline
    3 & Data Locality & Objects of similar types or objects used together frequently which are declared as \\ 
    & &individual objects\\
    \hline
   4 & Object Pool & Objects allocated on heap or similar objects created very frequently \\
    \hline
  \end{tabular}
  \label{rules}
\end{table*}


\subsection{Results}

\begin{figure}[ht!]
    \includegraphics[width=\columnwidth,height=6cm]{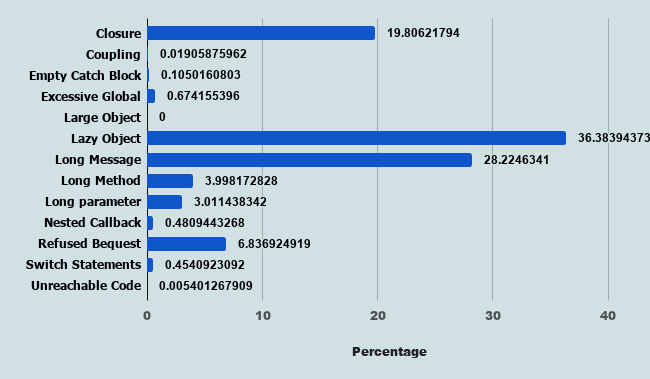}
    \caption{Distribution of code smells}
    \label{distribution}
\end{figure}

We did an empirical study on 361 \textit{JavaScript} games mined from Github to answer our first research question by evaluating all the mined games with the help of  \textit{JSNose}. Also, to comment on our second research question, we tried to manually figure out the violation of game programming patterns in 9 out of all the collected games.\newline
\textbf{RQ1: What is the distribution of existing code smells in context of games?}\newline
Distribution refers to the frequency and occurrence of different code smells detected by the tool. 
We analyzed that the majority of the code smells are primarily from 2 out of 13 code smells proposed by Fard \cite{fard2013jsnose}. The distribution is shown in Figure \ref{distribution}. We can infer from the figure that Lazy Object is the most frequent bad design practice during the development of web games with the occurrence of 36.3\% among all the smells. Following this, comes Long Message Chain with 28.22\%. The next most occurring is Closure smell with the occurrence of 19.80\% among all the smells. Following these are Refused Bequest, Long Method, Long Parameter List and others are negligible. Table \ref{jsnoseResults} depicts the statistical results we got by this empirical study. Here, we also showed the average occurrence of any code smells in all web games of our dataset, which again states that on an average approximately 1306 Lazy Object code smells existed in each web game of our dataset. Another criterion of Table shows the percentage of the games out of total games containing any specific smell. Out of all web games we took, 42.62\% of games contains Lazy Object code smell.
\newline
\newline
\textbf{RQ2: If game-specific code smells exists, how they are different from existing ones?}\newline
Manual study of 9 web games reveals that many game programming patterns that are desirable as proposed by \cite{nystrom2014game} are violated by different games. Since this is an initial study, we primarily focused on two broad classifications of game patterns, which are, Decoupling patterns and Optimization patterns. We observed that among Optimization patterns, Data Locality is the most violated pattern which is often overlooked by game developers. It was found in all the 9 games we studied manually.
This results in bad performance of games and affects the user game experience. Other violated patterns are shown in Table \ref{manualstudy} and our detailed report on the presence of violated patterns of 9 web games is present at the link \footnote{\url{http://bit.ly/GameSmellsJS}}.

\textbf{Violation of Component Decoupling}:
An instance of violation of Component Decoupling can be observed in \textit{Preloader.js} file of game \textit{Follow ME! A Simon Clone} \footnote{\url{https://github.com/gamedolphin/follow\_me\_javascript\_simon\_clone}} in line number 30-32. Different calls related to graphics, icons and audio component are there, but all are being called in the single function call. Thus, it may be refactored into different function calls, so that various components are decoupled.
\begin{scriptsize}
\codeJavascript
\begin{lstlisting}[frame=single]
this.load.atlas('spriteset', 'assets/spriteset.png', 'assets/spriteset.jsona');
        this.load.image('tweet','assets/twit.png');
        this.load.audio('sfx', ['assets/sfx.mp3','assets/sfx.ogg','assets/sfx.wav','assets/sfx.m4a']);
\end{lstlisting}
\end{scriptsize}

\textbf{Violation of Event Queue Decoupling}:
Whenever input events or events from the method are not stored in the central queue then, there occurs a violation of the event queue pattern. Below code snippet is taken from game \textit{Cube-Composer}\footnote{\url{https://github.com/sharkdp/cube-composer}} from the file \textit{Storage.js} from line number 1-11. We observe that the variable \textit{data} is getting input from the element which is not present in the central event queue of the game. 
To refactor the code smell, possibly we can make EventQueue class with methods such as \textit{Add()} and \textit{PublishEvents()}. The second method will be used at the end of the execution flow. Using this class will help in adding the events in the central queue and utilize them at the execution time. Also, nested return statements cause asynchronous behavior, hence delay in the occurrence of events.
\begin{scriptsize}
\codeJavascript
\begin{lstlisting}[frame=single]
exports.unsafeLoadGameState = function (just) {
    return function(nothing) {
        return function() {
            var data = localStorage.getItem('gameState');
            if (!data) {
                return nothing;
            }
            return just(JSON.parse(data));
        };
    };
};
\end{lstlisting}
\end{scriptsize}

\textbf{Violation of Data Locality}:
As an instance of Data Locality, the below code snippet is line 1-15 from \textit{Boot.js} file of the game \textit{JavaScript\_Snake}\footnote{\url{https://github.com/gamedolphin/javascript\_snake}}. The snippet follows the basic game variables declared in the form of \textit{JavaScript} object.
\begin{scriptsize}
\codeJavascript
\begin{lstlisting}[frame=single]
    var BasicGame = {
    //global variables
    timerDelay : 400,   //snake movement delay
    score : 0,          //current score
    highscore : null,   //object to store highscores
    currentMode : 'E',  //current play mode - easy/medium/hard
    trailno : 6,        //length of the trailing snake effect
    textList : null   //object to hold parsed game text
};
\end{lstlisting}
\end{scriptsize}
Since the objects will be accessed frequently, it can be refactored as an array object as shown below. Accessing array objects enables faster memory access and helps in taking advantage of data locality.
\begin{scriptsize}
\codeJavascript
\begin{lstlisting}[frame=single]
    timerDelay : 400,   //snake movement delay
    score : 0,          //current score
    highscore : null,   //object to store highscores
    currentMode : 'E',  //current play mode - easy/medium/hard
    trailno : 6,        //length of the trailing snake effect
    textList : null   //object to hold parsed game text
var BasicGame = new Array(timerDelay, score, highscore, currentMode, trailno, textList);

\end{lstlisting}
\end{scriptsize}

\textbf{Violation of Object Pool}:
An example of the violation of Object Pool can be shown in the below code snippet taken from line number 130-164 of \textit{keyboard.js} file of \textit{Astray}\footnote{\url{https://github.com/wwwtyro/Astray}}.
\begin{scriptsize}
\codeJavascript
\begin{lstlisting}[frame=single]
function queryActiveBindings() {
		var bindingStack = [];

		//loop through the key binding groups by number of keys.
		for(var keyCount = keyBindingGroups.length; keyCount > -1; keyCount -= 1) {
			if(keyBindingGroups[keyCount]) {
				var KeyBindingGroup = keyBindingGroups[keyCount];

				//loop through the key bindings of the same key length.
				for(var bindingIndex = 0; bindingIndex < KeyBindingGroup.length; bindingIndex += 1) {
					var binding = KeyBindingGroup[bindingIndex],

					//assume the binding is active till a required key is found to be unsatisfied
						keyBindingActive = true;

					//loop through each key required by the binding.
					for(var keyIndex = 0; keyIndex < binding.keys.length;  keyIndex += 1) {
						var key = binding.keys[keyIndex];

						//if the current key is not in the active keys array the mark the binding as inactive
						if(activeKeys.indexOf(key) < 0) {
							keyBindingActive = false;
						}
					}

					//if the key combo is still active then push it into the binding stack
					if(keyBindingActive) {
						bindingStack.push(binding);
					}}}}
		return bindingStack;
	}
\end{lstlisting}
\end{scriptsize}
We can observe that similar objects are created very frequently. This causes performance degradation and more processing time. Instead, an ObjectPool handler class can be created which will help in fetching and recycling the objects created in pool.

This manual study demonstrates the scope of the presence of additional code smells which have not been considered by existing code smell detection tools and not discussed in the literature in the domain of web games.
Thereafter, this observation shows that existing code smell detection tools are not sufficient to detect the bad smells of web games. There is a need for game-specific code smell detection tools. The violated game programming patterns are discussed in Section \ref{gamepatterns}.
\section{Related Work}
\label{relatedwork}

Various techniques and tools have been proposed to identify and detect code smells \cite{fard2013jsnose,palomba2017lightweight,shoenberger2017use}. Studies revolving around code smells deals mainly with its three divisions, which are, detection of code smells, evolution and impact of code smells, and the relation of software activities with code smells \cite{saboury2017empirical}.  Table \ref{tab:table1} represents some background work relevant to these three categories.\newline

\begin{minipage}{\linewidth}
\captionsetup{width=0.8\linewidth}
\captionof{table}{References Table}
  \centering
    \label{tab:table1}
    \begin{tabular}{|l|l|l|} 
    \hline
      \textbf{SN} & \textbf{Domain} & \textbf{References}\\
      \hline
      1 & Detection of code &  PMD\footnote{https://github.com/pmd/pmd}, FindBugs\footnote{http://bit.ly/2lmjsM2}, \\
       & smells & JLint \cite{artho2001finding}, SpIRIT \cite{vidal2016approach} \\
       & & Checkstyle\footnote{https://checkstyle.org/}\\ 
      2 & Evolution and Impact & \cite{olbrich2009evolution, olbrich2010all, chatzigeorgiou2010investigating, tahmid2016understanding}\\
      3 & Relationship to software  &  \cite{yamashita2013exploring, sjoberg2013quantifying, fontana2013investigating} \\
      & development process &  \\
      \hline
    \end{tabular}
\end{minipage}
\newline
The concern of code smells was identified by Fowler. He proposed that writing well factored program plays a major role in optimizing performance of any program \cite{fowler1997refactoring}. Following this, researchers proposed numerous code smells detecting rules and tools. Researchers also did empirical studies on patterns of bad smells and their detection. An initial study was done to understand bad smells in code and taxonomy was done to categorize similar code smells \cite{mantyla2003taxonomy}.
An empirical study done by Khomh et al. shows the relation of code smells to change-proneness by analyzing several releases of Azureus and Eclipse application \cite{khomh2009exploratory}. Johannes et al. conducted a large scale study on 1807 releases of 15 \textit{JavaScript} applications and detected 12 types of code smells \cite{johannes2019large}. They did a survival analysis of code containing code smells against codes without code smells until any fault occurs. Another empirical study was done to analyze \textit{JavaScript} callback usage as callbacks induces non-linear control flow of program and can cause asynchronous execution \cite{gallaba2015don}. A survey was done to understand the importance of code smells to developers and whether they really care about bad smells in code \cite{yamashita2013developers}. In a study, Fontana et al. tried to focus on some common problem faced by researchers for code smell detection and proposed a machine-learning based approach to deal with this problem \cite{fontana2013code}. Tahir et al. studied the discussions among developers on StackOverflow about code smells and anti-patterns \cite{tahir2018can}.

Code smells in android applications was also studied by researchers \cite{mannan2016understanding}, which states that some code smells occur frequently in android applications.
An android specific code smell detection tools, \textit{aDoctor} was proposed by Palomba et al. which detects 15 code smells \cite{palomba2017lightweight}.
To find the best threshold of metrics for code detection, Shoenberger et al. proposed a Genetic Programming approach to train the detection rules \cite{shoenberger2017use}. Studies have been done to identify and detect CSS code smells in web applications \cite{mesbah2012automated, araya2015creation}.

Besides empirical studies and surveys, researchers and developers have proposed numerous code smells detecting tools \cite{fernandes2016review}. A review study done on code smell detection tools reveals that majority of the existing tools are for Java programming language \cite{fernandes2016review}. They found 84 tools were proposed in literature for smell detection, out of which only 29 tools are available online. Fard et al. proposed 13 code smell detection technique along with a tool called \textit{JSNose} \cite{fard2013jsnose}. A tool called \textit{WebScent} \cite{nguyen2012detection} is proposed to detect embedded code smells but literature reveals it is unavailable online  \cite{fernandes2016review}. \textit{JSClassFinder} is a tool developed to find class-like structures in \textit{JavaScript} code \cite{silva2016jsclassfinder}. The tool is integration of a parser that takes the AST(Abstract Syntax Tree) of \textit{JavaScript} application along with Moose platform for visualization. To provide an interactive visualization of code smells for quick overview of code smells, researchers proposed a tool called \textit{Stench Blossom} \cite{murphy2010interactive}. It detects code smells and helps developers to understand the source of the smell. A meta-tool is proposed by combining different features of different code smell detector tools \cite{rutar2004comparison}. Many other tools were proposed for code quality improvement and software quality assurance for different programming languages \cite{tsantalis2008jdeodorant, vidal2015jspirit, yamashita2013extent}.

Although a lot of research work has been done on code quality and code smells, to the best of our knowledge, there exist very limited work on code smells in games. In a paper, Graham et al. discusses the code attributes required to develop quality games and thus illustrates the architectural desired practices with the help of a game  \cite{graham2006toward}. A related work on quality assurance of small games in a game jam setting is a recent survey of participants of Global Game Jam (GGJ), 48-hour hackathon, where researchers tried to study the effects of time pressure \cite{borg2019video}. They concluded that GGJ teams relies on ad hoc approach to develop and does face-to-face communication, and also these teams share contextual similarities to software startups.

\section{Threats to validity}
\label{discussion}
Since we are relying on \textit{JSNose} to detect the code smells in games, the accuracy of this tool decides the accuracy of our results.  \textit{JSNose} detects code smells in client-side code by using a metric-based approach which combines static and dynamic analysis. To mitigate this fact, we took help from existing literature which states that \textit{JSNose} has good precision score \cite{rasool2015review}. But the thresholds used to get these precision scores are empirical and may be erroneous. For reliability one may check \textit{JSNose} is available publicly.

One threat to the internal validity is our manual inspection of search of violation of game programming pattern. We aimed at investigating the violation of game programming patterns which have not been considered by existing code smell detection tools for \textit{JavaScript} language. To the best of our efforts and deep inspection, we believe that our observation is factual and gives the scope of the existence of game-specific code smells. But the validation and accuracy analysis performed by manual inspection can be incomplete and inaccurate.

The results we found are based on the 361 web games mined from Github. Thus, this result may vary with a large corpus of games. Although, we tried to mine as many web games as possible from Github, we found that many of the projects were private, inactive and some were filled with irrelevant codes \cite{kalliamvakou2014promises}. Also, since we did manual study only on 9 \textit{JavaScript} games, the results may vary for large scale manual study.
\section{Conclusion}
\label{conclusion}
In this paper, we tried to find out whether the existing code smells are relevant in the context of web games or not. The distribution of code smells in games and traditional software varies in the sense that performance and real-time response is the top requirement in case of games, which makes this domain unique \cite{murphy2014cowboys}. We observed that \textit{Lazy Object} is the most occurring code smell among all the 13 code smells proposed so far, following which comes the \textit{Long Message Chain} among most instances of occurrence. 

Our manual study done on 9 web games unfolded the concern that there is some violation of game programming pattern which needs to be paid attention in the domain of games. We proposed some of the code smells, which can be particular in the context of web games. The identification of these game-specific code smells can help game developers to be cautious while the development of games, which can avoid the risk of future maintainability, performance, and security.

\section{Future Work}
\label{future}
To further strengthen this research study on code smells in games, there are number of future directions. 
\begin{itemize}
    \item A systematic and efficient approach to discover game-specific code smells.
    \item We propose to create a catalog of game-specific code smells. We see a further need to extend our study to analyze more code smells in web games.
    \item Analyzing violation of game programming patterns manually, as done in this paper, is an effort-intensive task. We see a definite need to develop a code smell detection tool for the domain of games. 
    \item There is a vast amount of existing empirical research in mining software repositories focusing on APIs, pull requests, bug triaging and so on, which can be reproduced for the domain of games.
    \item We also intend to do a comparative study on presence and the effect of code smells in games and non-games.
\end{itemize}

\section*{Acknowledgements}
We would like to thank Vaishali Khanve for her valuable contributions to this project.
\balance
\bibliographystyle{IEEEannot}
\bibliography{games}

\begin{thebibliography}{10}
\providecommand{\url}[1]{#1}
\csname url@rmstyle\endcsname
\providecommand{\newblock}{\relax}
\providecommand{\bibinfo}[2]{#2}
\providecommand\BIBentrySTDinterwordspacing{\spaceskip=0pt\relax}
\providecommand\BIBentryALTinterwordstretchfactor{4}
\providecommand\BIBentryALTinterwordspacing{\spaceskip=\fontdimen2\font plus
\BIBentryALTinterwordstretchfactor\fontdimen3\font minus
  \fontdimen4\font\relax}
\providecommand\BIBforeignlanguage[2]{{%
\expandafter\ifx\csname l@#1\endcsname\relax
\typeout{** WARNING: IEEEtran.bst: No hyphenation pattern has been}%
\typeout{** loaded for the language `#1'. Using the pattern for}%
\typeout{** the default language instead.}%
\else
\language=\csname l@#1\endcsname
\fi
#2}}

\bibitem{abilio2015detecting}
R.~Ab{\'\i}lio, J.~Padilha, E.~Figueiredo, and H.~Costa, ``Detecting code
  smells in software product lines--an exploratory study,'' in \emph{2015 12th
  International Conference on Information Technology-New Generations}.\hskip
  1em plus 0.5em minus 0.4em\relax IEEE, 2015, pp. 433--438.


\bibitem{anderson2001effects}
C.~A. Anderson and B.~J. Bushman, ``Effects of violent video games on
  aggressive behavior, aggressive cognition, aggressive affect, physiological
  arousal, and prosocial behavior: A meta-analytic review of the scientific
  literature,'' \emph{Psychological science}, vol.~12, no.~5, pp. 353--359,
  2001.


\bibitem{araya2015creation}
C.~Araya, F.~Zoufaly, R.~Laplante, and O.~Calvo, ``Creation, generation,
  distribution and application of self-contained modifications to source
  code,'' Dec.~15 2015, uS Patent 9,213,541.


\bibitem{artho2001finding}
C.~Artho, ``Finding faults in multi-threaded programs,'' 2001.


\bibitem{bjork2006games}
S.~Bj{\"o}rk and J.~Holopainen, ``Games and design patterns,'' \emph{The game
  design reader}, pp. 410--437, 2006.


\bibitem{borg2019video}
M.~Borg, V.~Garousi, A.~Mahmoud, T.~Olsson, and O.~Stalberg, ``Video game
  development in a rush: A survey of the global game jam participants,''
  \emph{IEEE Transactions on Games}, 2019.


\bibitem{chatzigeorgiou2010investigating}
A.~Chatzigeorgiou and A.~Manakos, ``Investigating the evolution of bad smells
  in object-oriented code,'' in \emph{2010 Seventh International Conference on
  the Quality of Information and Communications Technology}.\hskip 1em plus
  0.5em minus 0.4em\relax IEEE, 2010, pp. 106--115.


\bibitem{crockford2008javascript}
D.~Crockford, \emph{JavaScript: The Good Parts: The Good Parts}.\hskip 1em plus
  0.5em minus 0.4em\relax " O'Reilly Media, Inc.", 2008.


\bibitem{fabricatore2007gameplay}
C.~Fabricatore, ``Gameplay and game mechanics: a key to quality in
  videogames,'' 2007.


\bibitem{fard2013jsnose}
A.~M. Fard and A.~Mesbah, ``Jsnose: Detecting javascript code smells,'' in
  \emph{2013 IEEE 13th International Working Conference on Source Code Analysis
  and Manipulation (SCAM)}.\hskip 1em plus 0.5em minus 0.4em\relax IEEE, 2013,
  pp. 116--125.


\bibitem{fernandes2016review}
E.~Fernandes, J.~Oliveira, G.~Vale, T.~Paiva, and E.~Figueiredo, ``A
  review-based comparative study of bad smell detection tools,'' in
  \emph{Proceedings of the 20th International Conference on Evaluation and
  Assessment in Software Engineering}.\hskip 1em plus 0.5em minus 0.4em\relax
  ACM, 2016, p.~18.


\bibitem{fontana2013investigating}
F.~A. Fontana, V.~Ferme, A.~Marino, B.~Walter, and P.~Martenka, ``Investigating
  the impact of code smells on system's quality: An empirical study on systems
  of different application domains,'' in \emph{2013 IEEE International
  Conference on Software Maintenance}.\hskip 1em plus 0.5em minus 0.4em\relax
  IEEE, 2013, pp. 260--269.


\bibitem{fontana2013code}
F.~A. Fontana, M.~Zanoni, A.~Marino, and M.~V. M{\"a}ntyl{\"a}, ``Code smell
  detection: Towards a machine learning-based approach,'' in \emph{2013 IEEE
  International Conference on Software Maintenance}.\hskip 1em plus 0.5em minus
  0.4em\relax IEEE, 2013, pp. 396--399.


\bibitem{fowler1997refactoring}
M.~Fowler, ``Refactoring: Improving the design of existing code,'' in
  \emph{11th European Conference. Jyv{\"a}skyl{\"a}, Finland}, 1997.


\bibitem{gallaba2015don}
K.~Gallaba, A.~Mesbah, and I.~Beschastnikh, ``Don't call us, we'll call you:
  Characterizing callbacks in javascript,'' in \emph{2015 ACM/IEEE
  International Symposium on Empirical Software Engineering and Measurement
  (ESEM)}.\hskip 1em plus 0.5em minus 0.4em\relax IEEE, 2015, pp. 1--10.


\bibitem{ghafari2017security}
M.~Ghafari, P.~Gadient, and O.~Nierstrasz, ``Security smells in android,'' in
  \emph{2017 IEEE 17Th international working conference on source code analysis
  and manipulation (SCAM)}.\hskip 1em plus 0.5em minus 0.4em\relax IEEE, 2017,
  pp. 121--130.


\bibitem{giessen2015serious}
H.~W. Giessen, ``Serious games effects: an overview,'' \emph{Procedia-Social
  and Behavioral Sciences}, vol. 174, pp. 2240--2244, 2015.


\bibitem{graham2006toward}
T.~N. Graham and W.~Roberts, ``Toward quality-driven development of 3d computer
  games,'' in \emph{International Workshop on Design, Specification, and
  Verification of Interactive Systems}.\hskip 1em plus 0.5em minus 0.4em\relax
  Springer, 2006, pp. 248--261.


\bibitem{guo2010domain}
Y.~Guo, C.~Seaman, N.~Zazworka, and F.~Shull, ``Domain-specific tailoring of
  code smells: an empirical study,'' in \emph{Proceedings of the 32nd ACM/IEEE
  International Conference on Software Engineering-Volume 2}.\hskip 1em plus
  0.5em minus 0.4em\relax ACM, 2010, pp. 167--170.


\bibitem{hecht2016empirical}
G.~Hecht, N.~Moha, and R.~Rouvoy, ``An empirical study of the performance
  impacts of android code smells,'' in \emph{Proceedings of the International
  Conference on Mobile Software Engineering and Systems}.\hskip 1em plus 0.5em
  minus 0.4em\relax ACM, 2016, pp. 59--69.


\bibitem{hermans2012detecting}
F.~Hermans, M.~Pinzger, and A.~van Deursen, ``Detecting code smells in
  spreadsheet formulas,'' in \emph{2012 28th IEEE International Conference on
  Software Maintenance (ICSM)}.\hskip 1em plus 0.5em minus 0.4em\relax IEEE,
  2012, pp. 409--418.


\bibitem{johannes2019large}
D.~Johannes, F.~Khomh, and G.~Antoniol, ``A large-scale empirical study of code
  smells in javascript projects,'' \emph{Software Quality Journal}, pp. 1--44,
  2019.


\bibitem{kalliamvakou2014promises}
E.~Kalliamvakou, G.~Gousios, K.~Blincoe, L.~Singer, D.~M. German, and
  D.~Damian, ``The promises and perils of mining github,'' in \emph{Proceedings
  of the 11th working conference on mining software repositories}.\hskip 1em
  plus 0.5em minus 0.4em\relax ACM, 2014, pp. 92--101.


\bibitem{kasurinen2013game}
J.~Kasurinen, J.-P. Strand{\'e}n, and K.~Smolander, ``What do game developers
  expect from development and design tools?'' in \emph{Proceedings of the 17th
  International Conference on Evaluation and Assessment in Software
  Engineering}.\hskip 1em plus 0.5em minus 0.4em\relax ACM, 2013, pp. 36--41.


\bibitem{kerievsky2005refactoring}
J.~Kerievsky, \emph{Refactoring to patterns}.\hskip 1em plus 0.5em minus
  0.4em\relax Pearson Deutschland GmbH, 2005.


\bibitem{khanve2019existing}
V.~Khanve, ``Are existing code smells relevant in web games? an empirical
  study,'' in \emph{Proceedings of the 2019 27th ACM Joint Meeting on European
  Software Engineering Conference and Symposium on the Foundations of Software
  Engineering}.\hskip 1em plus 0.5em minus 0.4em\relax ACM, 2019, pp.
  1241--1243.


\bibitem{khomh2009exploratory}
F.~Khomh, M.~Di~Penta, and Y.-G. Gueheneuc, ``An exploratory study of the
  impact of code smells on software change-proneness,'' in \emph{2009 16th
  Working Conference on Reverse Engineering}.\hskip 1em plus 0.5em minus
  0.4em\relax IEEE, 2009, pp. 75--84.


\bibitem{klimmt2009player}
C.~Klimmt, C.~Blake, D.~Hefner, P.~Vorderer, and C.~Roth, ``Player performance,
  satisfaction, and video game enjoyment,'' in \emph{International Conference
  on Entertainment Computing}.\hskip 1em plus 0.5em minus 0.4em\relax Springer,
  2009, pp. 1--12.


\bibitem{kostanjevec2017preliminary}
D.~Kostanjevec, M.~Pusnik, M.~Hericko, B.~Sumak, G.~Rakic, and Z.~Budimac, ``A
  preliminary empirical exploration of quality measurement for javascript
  solutions.'' in \emph{SQAMIA}, 2017.


\bibitem{lee2006we}
K.~M. Lee and W.~Peng, ``What do we know about social and psychological effects
  of computer games? a comprehensive review of the current literature,''
  \emph{Playing video games: Motives, responses, and consequences}, pp.
  327--345, 2006.


\bibitem{mannan2016understanding}
U.~A. Mannan, I.~Ahmed, R.~A.~M. Almurshed, D.~Dig, and C.~Jensen,
  ``Understanding code smells in android applications,'' in \emph{2016 IEEE/ACM
  International Conference on Mobile Software Engineering and Systems
  (MOBILESoft)}.\hskip 1em plus 0.5em minus 0.4em\relax IEEE, 2016, pp.
  225--236.


\bibitem{mantyla2003taxonomy}
M.~Mantyla, J.~Vanhanen, and C.~Lassenius, ``A taxonomy and an initial
  empirical study of bad smells in code,'' in \emph{International Conference on
  Software Maintenance, 2003. ICSM 2003. Proceedings.}\hskip 1em plus 0.5em
  minus 0.4em\relax IEEE, 2003, pp. 381--384.


\bibitem{marchand2013value}
A.~Marchand and T.~Hennig-Thurau, ``Value creation in the video game industry:
  Industry economics, consumer benefits, and research opportunities,''
  \emph{Journal of Interactive Marketing}, vol.~27, no.~3, pp. 141--157, 2013.


\bibitem{mesbah2012automated}
A.~Mesbah and S.~Mirshokraie, ``Automated analysis of css rules to support
  style maintenance,'' in \emph{Proceedings of the 34th International
  Conference on Software Engineering}.\hskip 1em plus 0.5em minus 0.4em\relax
  IEEE Press, 2012, pp. 408--418.


\bibitem{murphy2010interactive}
E.~Murphy-Hill and A.~P. Black, ``An interactive ambient visualization for code
  smells,'' in \emph{Proceedings of the 5th international symposium on Software
  visualization}.\hskip 1em plus 0.5em minus 0.4em\relax ACM, 2010, pp. 5--14.


\bibitem{murphy2014cowboys}
E.~Murphy-Hill, T.~Zimmermann, and N.~Nagappan, ``Cowboys, ankle sprains, and
  keepers of quality: How is video game development different from software
  development?'' in \emph{Proceedings of the 36th International Conference on
  Software Engineering}.\hskip 1em plus 0.5em minus 0.4em\relax ACM, 2014, pp.
  1--11.


\bibitem{nguyen2012detection}
H.~V. Nguyen, H.~A. Nguyen, T.~T. Nguyen, A.~T. Nguyen, and T.~N. Nguyen,
  ``Detection of embedded code smells in dynamic web applications,'' in
  \emph{2012 Proceedings of the 27th IEEE/ACM International Conference on
  Automated Software Engineering}.\hskip 1em plus 0.5em minus 0.4em\relax IEEE,
  2012, pp. 282--285.


\bibitem{nystrom2014game}
R.~Nystrom, \emph{Game programming patterns}.\hskip 1em plus 0.5em minus
  0.4em\relax Genever Benning, 2014.


\bibitem{olbrich2009evolution}
S.~Olbrich, D.~S. Cruzes, V.~Basili, and N.~Zazworka, ``The evolution and
  impact of code smells: A case study of two open source systems,'' in
  \emph{2009 3rd international symposium on empirical software engineering and
  measurement}.\hskip 1em plus 0.5em minus 0.4em\relax IEEE, 2009, pp.
  390--400.


\bibitem{olbrich2010all}
S.~M. Olbrich, D.~S. Cruzes, and D.~I. Sj{\o}berg, ``Are all code smells
  harmful? a study of god classes and brain classes in the evolution of three
  open source systems,'' in \emph{2010 IEEE International Conference on
  Software Maintenance}.\hskip 1em plus 0.5em minus 0.4em\relax IEEE, 2010, pp.
  1--10.


\bibitem{palomba2017lightweight}
F.~Palomba, D.~Di~Nucci, A.~Panichella, A.~Zaidman, and A.~De~Lucia,
  ``Lightweight detection of android-specific code smells: The adoctor
  project,'' in \emph{2017 IEEE 24th International Conference on Software
  Analysis, Evolution and Reengineering (SANER)}.\hskip 1em plus 0.5em minus
  0.4em\relax IEEE, 2017, pp. 487--491.


\bibitem{rasool2015review}
G.~Rasool and Z.~Arshad, ``A review of code smell mining techniques,''
  \emph{Journal of Software: Evolution and Process}, vol.~27, no.~11, pp.
  867--895, 2015.


\bibitem{ray2014large}
B.~Ray, D.~Posnett, V.~Filkov, and P.~Devanbu, ``A large scale study of
  programming languages and code quality in github,'' in \emph{Proceedings of
  the 22nd ACM SIGSOFT International Symposium on Foundations of Software
  Engineering}.\hskip 1em plus 0.5em minus 0.4em\relax ACM, 2014, pp. 155--165.


\bibitem{rutar2004comparison}
N.~Rutar, C.~B. Almazan, and J.~S. Foster, ``A comparison of bug finding tools
  for java,'' in \emph{15th International Symposium on Software Reliability
  Engineering}.\hskip 1em plus 0.5em minus 0.4em\relax IEEE, 2004, pp.
  245--256.


\bibitem{saboury2017empirical}
A.~Saboury, P.~Musavi, F.~Khomh, and G.~Antoniol, ``An empirical study of code
  smells in javascript projects,'' in \emph{2017 IEEE 24th international
  conference on software analysis, evolution and reengineering (SANER)}.\hskip
  1em plus 0.5em minus 0.4em\relax IEEE, 2017, pp. 294--305.


\bibitem{sharma2018survey}
T.~Sharma and D.~Spinellis, ``A survey on software smells,'' \emph{Journal of
  Systems and Software}, vol. 138, pp. 158--173, 2018.


\bibitem{shoenberger2017use}
I.~Shoenberger, M.~W. Mkaouer, and M.~Kessentini, ``On the use of smelly
  examples to detect code smells in javascript,'' in \emph{European Conference
  on the Applications of Evolutionary Computation}.\hskip 1em plus 0.5em minus
  0.4em\relax Springer, 2017, pp. 20--34.


\bibitem{silva2016jsclassfinder}
L.~H. Silva, D.~Hovadick, M.~T. Valente, A.~Bergel, N.~Anquetil, and A.~Etien,
  ``Jsclassfinder: A tool to detect class-like structures in javascript,''
  \emph{arXiv preprint arXiv:1602.05891}, 2016.


\bibitem{sjoberg2013quantifying}
D.~I. Sj{\o}berg, A.~Yamashita, B.~C. Anda, A.~Mockus, and T.~Dyb{\aa},
  ``Quantifying the effect of code smells on maintenance effort,'' \emph{IEEE
  Transactions on Software Engineering}, vol.~39, no.~8, pp. 1144--1156, 2013.


\bibitem{stamelos2002code}
I.~Stamelos, L.~Angelis, A.~Oikonomou, and G.~L. Bleris, ``Code quality
  analysis in open source software development,'' \emph{Information Systems
  Journal}, vol.~12, no.~1, pp. 43--60, 2002.


\bibitem{tahir2018can}
A.~Tahir, A.~Yamashita, S.~Licorish, J.~Dietrich, and S.~Counsell, ``Can you
  tell me if it smells?: A study on how developers discuss code smells and
  anti-patterns in stack overflow,'' in \emph{Proceedings of the 22nd
  International Conference on Evaluation and Assessment in Software Engineering
  2018}.\hskip 1em plus 0.5em minus 0.4em\relax ACM, 2018, pp. 68--78.


\bibitem{tahmid2016understanding}
A.~Tahmid, N.~Nahar, and K.~Sakib, ``Understanding the evolution of code smells
  by observing code smell clusters,'' in \emph{2016 IEEE 23rd International
  Conference on Software Analysis, Evolution, and Reengineering (SANER)},
  vol.~4.\hskip 1em plus 0.5em minus 0.4em\relax IEEE, 2016, pp. 8--11.


\bibitem{tsantalis2008jdeodorant}
N.~Tsantalis, T.~Chaikalis, and A.~Chatzigeorgiou, ``Jdeodorant: Identification
  and removal of type-checking bad smells,'' in \emph{2008 12th European
  Conference on Software Maintenance and Reengineering}.\hskip 1em plus 0.5em
  minus 0.4em\relax IEEE, 2008, pp. 329--331.


\bibitem{van2002java}
E.~Van~Emden and L.~Moonen, ``Java quality assurance by detecting code
  smells,'' in \emph{Ninth Working Conference on Reverse Engineering, 2002.
  Proceedings.}\hskip 1em plus 0.5em minus 0.4em\relax IEEE, 2002, pp. 97--106.


\bibitem{vidal2015jspirit}
S.~Vidal, H.~Vazquez, J.~A. Diaz-Pace, C.~Marcos, A.~Garcia, and W.~Oizumi,
  ``Jspirit: a flexible tool for the analysis of code smells,'' in \emph{2015
  34th International Conference of the Chilean Computer Science Society
  (SCCC)}.\hskip 1em plus 0.5em minus 0.4em\relax IEEE, 2015, pp. 1--6.


\bibitem{vidal2016approach}
S.~A. Vidal, C.~Marcos, and J.~A. D{\'\i}az-Pace, ``An approach to prioritize
  code smells for refactoring,'' \emph{Automated Software Engineering},
  vol.~23, no.~3, pp. 501--532, 2016.


\bibitem{yamashita2012code}
A.~Yamashita and L.~Moonen, ``Do code smells reflect important maintainability
  aspects?'' in \emph{2012 28th IEEE international conference on software
  maintenance (ICSM)}.\hskip 1em plus 0.5em minus 0.4em\relax IEEE, 2012, pp.
  306--315.


\bibitem{yamashita2013developers}
------, ``Do developers care about code smells? an exploratory survey,'' in
  \emph{2013 20th Working Conference on Reverse Engineering (WCRE)}.\hskip 1em
  plus 0.5em minus 0.4em\relax IEEE, 2013, pp. 242--251.


\bibitem{yamashita2013exploring}
------, ``Exploring the impact of inter-smell relations on software
  maintainability: An empirical study,'' in \emph{Proceedings of the 2013
  International Conference on Software Engineering}.\hskip 1em plus 0.5em minus
  0.4em\relax IEEE Press, 2013, pp. 682--691.


\bibitem{yamashita2013extent}
------, ``To what extent can maintenance problems be predicted by code smell
  detection?--an empirical study,'' \emph{Information and Software Technology},
  vol.~55, no.~12, pp. 2223--2242, 2013.


\end{thebibliography}
\end{document}